\documentclass[aps,prb,reprint]{revtex4-2}

\usepackage{graphicx}
\usepackage{amsmath}

\usepackage{tikz-feynman}
\tikzfeynmanset{compat=1.1.0}
\usetikzlibrary{shapes,arrows,positioning,automata,backgrounds,calc,er,patterns}

\newcommand{\bvec}[1]{\mathbf{#1}}

\begin{document}


\title{Interplay of exchange and spin-conserving scattering processes in a ferromagnetic two-sublattice system}
\author{A. Tavakoli}
\author{K. Leckron}
\author{H. C. Schneider}
\affiliation{Physics Department and Research Center OPTIMAS, RPTU University Kaiserslautern-Landau, 67663 Kaiserslautern, Germany}

\date{\today}

\begin{abstract}
In magnetic alloys or hybrid systems formed by molecular magnets coupled to a magnetic substrate, the exchange interaction leads to fast spin dynamics after excitation. We investigate the electronic charge and spin dynamics due to the exchange interaction in a two-sublattice system with antiferromagnetic coupling. We employ a simplified  model for a ferrimagnetic alloy as a coupled system of itinerant and localized electron states together with an exchange coupling between the two. For the itinerant system we include electron-electron Coulomb scattering and electron-phonon scattering. We study numerically the heat-induced ultrafast magnetization dynamics due to the interplay of exchange scattering and spin-independent scattering processes and discuss different scenarios for the demagnetization and relaxation dynamics of the sublattices. Our results highlight the impact of spin-conserving electron-electron scattering processes on the exchange-driven spin dynamics on ultrashort timescales. 
\end{abstract}
\maketitle

\section{Introduction\label{sec:introduction}}

Exchange interactions in magnetic alloys or hybrid systems formed by molecular magnets coupled to a magnetic substrate give rise to magnetization dynamics on ultrashort timescales following an optical excitation with an ultrashort laser pulse that can be investigated by element-specific experimental techniques, thus allowing one to study exchange interactions on their intrinsic timescale. The first experiments involved ferrimagnets with antiferromagnetic coupling such as GdFeCo, which may lead to a distinctive transient ferromagnetic-like state, but also leads to observable effects in ferromagnetic coupling alloys such as NiFe and FeCo. For the case of ferromagnetic coupling, the important signature observed in experiments is the delay between the element-specific magnetization traces, sometimes even an increase in one of the element-specific magnetization signals, and slow remagnetization dynamics~\cite{mathias_probing_2012,radu_ultrafast_2015,von_korff_schmising_element-specific_2020,yao_distinct_2020}.
The delay and also the element-specific increase has been related to an optically induced spin transfer (OISTR) mechanism, which describes a change of the element-specific contribution to the magnetic moment due to laser-pulse excitation, as implemented in time-dependent DFT~\cite{dewhurst_laser-induced_2018}. The OISTR effect arises from the coupling of the coherent electromagnetic field to the microscopic charge/current densities in the material; it therefore typically needs very high fluences~\cite{Mrudul}. Generally speaking, the OISTR effect may be called a ``primary effect''~\cite{Stiehl} as it imprints the details of the optical field in the excited states of the magnetic material directly. 

We distinguish such primary processes from secondary processes. The latter also change the (sublattice) magnetization following optical-pulse excitation, but are due to electronic interaction mechanisms that do not directly involve the coupling to the optical field. Recent studies have shown that secondary effects such as scattering and transport may also lead to element-specific dynamics of ferromagnetically coupled alloys that are similar to signatures of OISTR processes. For instance, experiments on FeNi showed element-specific dynamics after injection of hot electrons without a direct influence of a coherent driving field~\cite{Schmising}, which rules out OISTR contributions. Using a two-sublattice system with ferromagnetic exchange as model for a ferromagnetic alloy, we have presented theoretical evidence that incoherent exchange scattering could produce effects qualitatively similar to those of OISTR~in Ref.~\cite{Leckron2022}. The excitation condition analyzed in our earlier paper was an instantaneous heating of the electron system, which is an idealized model of the energy deposited in the magnetic system by an ultrashort optical pulse.

In the present paper we extend the study of exchange scattering processes in alloys presented in Ref.~\cite{Leckron2022}, where we employed a quasi-equilibrium assumption for the excitation. Our goal is twofold. First, we analyze initial conditions that mimic the electronic non-equilibrium created by optical fields and, second, we take into account electron-electron scattering. Depending on the photon energy and matrix elements, the optical field is only resonant with electronic transitions in a small region of the Brillouin zone, so that they generally create electronic distributions that are far from equilibrium. On very short timescales these electronic distributions are very different from a quasi-equilibrium. For this reason, scattering processes that act on the time scales set by a typically 50 fs pulse, in particular, electron-electron scattering processes, have to be taken into account. The second goal is to assess the interplay of these direct electron-electron scattering processes with exchange scattering processes between the sublattices. 

In our approach we directly describe the competing interaction mechanisms on ultrafast timescales in an electronic band picture. We therefore avoid a bath assumption for the electronic system in localized-spin models, which are often used to describe the collective spin dynamics in multi-sublattice systems.   

This paper is organized as follows. We first briefly introduce our model setup and the dynamical equations for the distribution functions that lead to the different scattering contribution. From these spin-dependent distribution functions, we obtain ensemble-averaged single-particle quantities such as ensemble spin expectation value/magnetization and total energy. We then present and compare results with and without non-exchange electron-electron scattering in a two-sublattice system with \emph{ferromagnetic} coupling.
\section{\label{sec:theory}Theoretical approach}

Figure~\ref{fig:bands} shows a sketch of the electronic band structure and the included interaction mechanisms. Our model band structure consists of exchange-coupled  sublattices that describe delocalized (sublattice A) and localized (sublattice B) electron states. This simplified band structure was introduced for \emph{antiferro}magnetically coupled sublattices as in GdFeCo~\cite{Baral2015} and adapted for a ferromagnetic inter-sublattice coupling in Ref.~\cite{Leckron2022} in order to capture some important features of ferro\-magnetic alloys. A major difference to Ref.~\cite{Leckron2022} is the inclusion of electron-electron scattering processes. Because of the large numerical effort of electron-electron scattering for a system with multiple itinerant bands, we retain the model setup of Ref.~\cite{Leckron2022}, with two itinerant bands as shown as the light blue and dark blue lines in Fig.~\ref{fig:bands}. We use a parabolic band structure for sublattice A (right side) and flat bands for sublattice B (left side) as extreme cases of element-specific differences in the sublattice band structure. The different scattering mechanisms are also indicated: Exchange scattering processes involve opposite spin-flips in sublattice A and sublattice B. In the itinerant subsystem A we also include``Elliott-Yafet'' scattering processes that flip a spin in subsystem A, and spin-conserving electron-electron scattering processes. Therefore, we will first give a brief summary of the main features of the model system treated in Ref.~\cite{Leckron2022}. 

\subsection{\label{sec:model}Model Hamiltonian}

\begin{figure}[t]
   \centering
\includegraphics[width=\columnwidth]{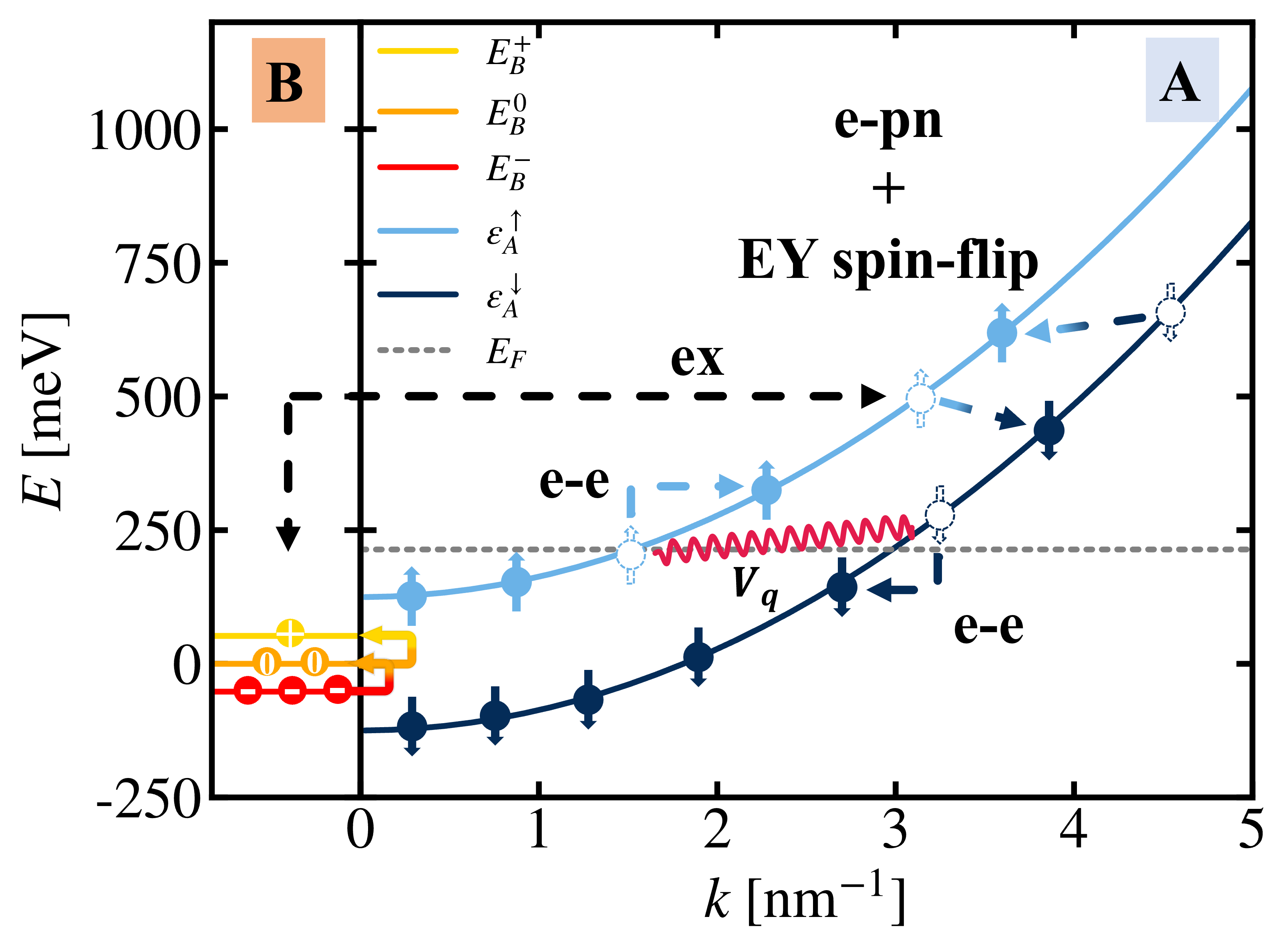}
	\caption{Sketch of the band structure of sublattice A and sublattice B including the corresponding scattering mechanisms. The minority and majority bands in sublattice A have opposite spin orientations, as depicted by the arrows. $E_F$ denotes the Fermi energy level in sublattice A. The parameters used here are $U_{\mathrm{eff}}=-100$ meV and $J=-200$ meV. The electron density in both sublattices is $n_A = n_B= 1$ $\mathrm{nm}^{-2}$. }
	\label{fig:bands}	
\end{figure}

We discuss here the Hamiltonian for the two-sublattice model with exchange coupling that defines the single-particle states and energies illustrated in Fig.~\ref{fig:bands}. The  contributions for the itinerant sublattice A and the localized (flat-band) sublattice B are given by $\hat{H_A}\text{(k)}=\hat{H}_\text{kin}+\hat{H}_\text{Stoner}+\hat{H}_\text{ex}$ and $\hat{H_B}=\hat{H}_\text{ex}$, respectively.
The contributions for sublattice A are the kinetic part $\hat{H}_{\text{kin}} = \frac{\hbar^2 k^2}{2m^*}$, where $k$ is the wave number of the vacuum electron mass $m^*=m_0$ is the effective mass,  the spin-dependent mean-field (Stoner) contribution $\hat{H}_{\text{Stoner}}=\frac{4}{3}U_{\mathrm{eff}}\hat{s_z}\langle \hat{s_z} \rangle$ and an exchange interaction Hamiltonian $H_{\text{ex}}=-J[\langle \hat{S_{z}}\rangle \hat{s}_z +\langle \hat{s_{z}} \rangle \hat{S}_z ]$. We have defined here an effective on-site interaction $U_{\mathrm{eff}}$ (Stoner parameter) and the spin operators of sublattice A $\hat{s}_z$, and B $\hat{S}_z$.  The exchange contribution connects sublattice A and sublattice B, while the on-site (Stoner) contribution acts only on itinerant electrons in sublattice A. The effective mass Hamiltonian leads to the parabolic band shape and $\hat{H}_{\text{Stoner}}$ introduces a spin-dependent splitting of the sublattice A band structure. Finally, $H_{\text{ex}}$ modifies this gap in sublattice A and is also responsible for the splitting of the flat bands in sublattice B. In this context, we treat the effective Coulomb potential $U_{\mathrm{eff}}$ and the exchange constant $J$ as material parameters.

The eigenenergies in sublattice A are labeled by $\nu=\uparrow$ and $\downarrow$, corresponding to minority and majority bands, respectively. They are given by 
\begin{equation}
    \epsilon_{\bvec{k}}^{\nu}= \frac{\hbar^2k^2}{2 m^*} 
    \begin{cases} 
        + \Delta, &\nu=\uparrow \\
        - \Delta, &\nu=\downarrow
    \end{cases}
    \label{eq:itinen}
\end{equation}
where the splitting of the itinerant states is determined by
\begin{equation}
	\Delta \equiv - \frac{2}{3} U_{\mathrm{eff}}\langle \hat{s}_{z} \rangle - \frac{1}{2} J \langle \hat{S}_{z} \rangle. 
	\label{eq:itindelta}
\end{equation}
The spin-1 sublattice B states are labeled by a single quantum number $\lambda$ and have the energies 
\begin{equation}
    E_{\lambda} =
    \begin{cases}
      +J\langle \hat{s_{z}}\rangle,  & \lambda = +1\\ 
      \quad 0, & \lambda = 0 \\ 
      -J\langle \hat{s_{z}}\rangle, &\lambda = -1
    \end{cases}
    \label{eq:loken}
\end{equation}
The eigenenergies and states introduced here depend on spin expectation values that change with time when the system is excited. They form adiabatic, instantaneous eigenstates of the exchange coupled system which enter into the calculation of the ultrafast spin-dependent electronic dynamics presented in the next section. 

\subsection{Equation of motion for electronic occupation numbers}

The basic dynamical quantities in our approach are the time-dependent occupations of the single-particle states introduced in the last section. If we define Heisenberg-picture creation and destruction operators $\hat{c}_{\bvec{k}\nu}^{\dagger}$ and $ \hat{c}_{\bvec{k}\nu}$ for the itinerant electron states states in subsystem A, the dynamical distribution function is given by $n_{\bvec{k}\nu}(t)= \langle \hat{c}_{\bvec{k}\nu}^{\dagger} \hat{c}_{\bvec{k}\nu} \rangle$ of the states. The corresponding quantities for the localized states are $\hat{c}_{\lambda}^{\dagger}$ and $ \hat{c}_{\lambda}$, leading to the dynamical occupations $n_\lambda (t) =\hat{c}_{\lambda}^{\dagger}\hat{c}_{\lambda} $ of the three states in subsystem B. As already mentioned, the states in subsystem B can be viewed as the limiting case of a flat band with vanishing $k$-dispersion. The derivation for all considered scattering mechanisms is described in detail in Refs.~\cite{Baral2015, Leckron2022, Vollmar2023}.

We present next only the final Boltzmann-like scattering integrals for the electron-electron, electron-phonon, and exchange interactions. For subsystem A these read
\begin{equation}
\frac{\partial}{\partial t} n^{\nu}_{\bvec{k}} = \frac{\partial}{\partial t} n^{\nu }_{\bvec{k}}\Bigg|_{\mathrm{e-e}} 
+ \frac{\partial}{\partial t} n^{\nu }_{\bvec{k}}\Bigg|_{\mathrm{e-pn}}
+ \frac{\partial}{\partial t} n^{\nu}_{\bvec{k}}\Bigg|_{\mathrm{ex}} 
+ \frac{\partial}{\partial t} n^{\nu}_{\bvec{k}}\Bigg|_{\mathrm{sr}} 
\end{equation}
with the contributions
\begin{widetext}
\begin{align}
 \frac{\partial}{\partial t} n^{\nu}_{\bvec{k}}\Bigg|_{\mathrm{e-e}} = &\frac{2}{\hbar} \sum_{\bvec{l} \bvec{q}} \sum_{\nu'}  \zeta \Biggl( \left| V_{\bvec{k}, \bvec{l}, \bvec{q}}^{\nu \nu'}\right|^{2}
		  \frac{n_{\bvec{k}+\bvec{q}}^{\nu} n_{\bvec{l}}^{\nu'}  \big( 1 - n_{\bvec{l}+\bvec{q}}^{\nu'} \big) \big( 1 - n_{\bvec{k}}^{\nu} \big) 
   - n_{\bvec{l}+\bvec{q}}^{\nu'}  n_{\bvec{k}}^{\nu} \big( 1 - n_{\bvec{l}}^{\nu'} \big) \big( 1 - n_{\bvec{k}+\bvec{q}}^{\nu} \big) }{ \epsilon_{\bvec{k}+\bvec{q}}^{\nu} - \epsilon_{\bvec{k}}^{\nu} + \epsilon_{\bvec{l}}^{\nu'}  - \epsilon_{\bvec{l}+\bvec{q}}^{\nu'} -i\hbar\Gamma} \Biggr), 
   \label{eq:EOM-e-e}\\
 \frac{\partial}{\partial t} n^{\nu}_{\bvec{k}}\Bigg|_{\mathrm{e-pn}} = &\frac{2}{\hbar} \sum_{k_1 \nu_1} \zeta \Biggl( \left|g_{\bvec{k} \nu , \bvec{k}_1 \nu_1}\right|^2
		  \frac{\left( 1+ N_{\bvec{k}_1 -\bvec{k}} \right) \left( 1 - n_{\bvec{k}}^{\nu} \right) n_{\bvec{k}_1}^{\nu_1}  - N_{\bvec{k}_1-\bvec{k}} \left( 1 -n_{\bvec{k}_1}^{\nu_1} \right) n_{\bvec{k}}^{\nu}}{\epsilon_{\bvec{k}}^{\nu} - \epsilon_{\bvec{k}_1}^{\nu_1} + \hbar\omega_{\bvec{k}_1-\bvec{k}} + i\hbar\Gamma} \Biggr),
   \label{eq:EOM-e-pn} \\
 \frac{\partial}{\partial t} n^{\nu}_{\bvec{k}}\Bigg|_{\mathrm{ex}} = &\frac{2}{\hbar}  \sum_{\bvec{k}' \nu'}\sum_{\lambda\lambda'} \zeta \Biggl(  \left|W_{\bvec{k} \nu, \bvec{k'} \nu'}^{\lambda \lambda'}\right|^2
		  \frac{n_{\bvec{k}'}^{\nu'}n_{B}^{\lambda'}\left(1-n_{\bvec{k}}^{\nu}\right)\left(1-n_{B}^{\lambda}\right)
	-n_{\bvec{k}}^{\nu}n_{B}^{\lambda}\left(1-n_{\bvec{k}'}^{\nu'}\right)\left(1-n_{B}^{\lambda'}\right) }{ E_{\lambda'} -  \epsilon_{\bvec{k}}^{\nu} + \epsilon_{\bvec{k}'}^{\nu'} -E_{\lambda} -i\hbar\Gamma} \Biggr),
   \label{eq:EOM-ex-ss1}
\end{align}
Further, for the localized states we have
\begin{equation}
 \frac{\partial}{\partial t} n^{\lambda}_{B}\Bigg|_{\mathrm{ex}} = \frac{2}{\hbar} \sum_{\bvec{k}\nu}\sum_{\bvec{k'}\nu'}\sum_{\lambda'} \zeta \Biggl(  \left|W_{\bvec{k} \nu, \bvec{k'} \nu'}^{\lambda \lambda'}\right|^2 \frac{n_{\bvec{k}'}^{\nu'}n_{B}^{\lambda'}\left(1-n_{\bvec{k}}^{\nu}\right)\left(1-n_{B}^{\lambda}\right)
	-n_{\bvec{k}}^{\nu}n_{B}^{\lambda}\left(1-n_{\bvec{k}'}^{\nu'}\right)\left(1-n_{B}^{\lambda'}\right) }{ E_{\lambda'} -  \epsilon_{\bvec{k}}^{\nu} + \epsilon_{\bvec{k}'}^{\nu'} -E_{\lambda} -i\hbar\Gamma} \Biggr),
   \label{eq:EOM-ex-ss2}
\end{equation}
\end{widetext}
In the equations above, the energies and states at each time depend on the instantaneous spin expectation values of the two subsystems.
Eq.~\eqref{eq:EOM-e-e} is the EOM for spin conserving electron-electron (e-e) scattering in which 
$ V_{\bvec{k}, \bvec{l}, \bvec{q}}^{\nu \nu'}=V^{\text{scr}}_q\langle \nu, \bvec{k}|\nu,\bvec{k}+\bvec{q}\rangle \langle \nu', \bvec{l}+\bvec{q}|\nu',\bvec{l}\rangle$ is the Coulomb-matrix element that depends on the momentum $q$ transferred in a transition from initial electronic states $|\nu,\bvec{k}\rangle,|\nu',\bvec{l}+\bvec{q}\rangle$ to final states $|\nu,\bvec{k}+\bvec{q}\rangle,|\nu',\bvec{l}\rangle$, as shown by the diagram in Fig.~\ref{fig:e-e-sketch}. Also, this interaction contribution is sketched as \textbf{e-e} in Fig.~\ref{fig:bands}. $V^{\text{scr}}_q \propto \frac{1}{\kappa^2 + q^2}$ denotes a screened Coulomb potential in 2-D where $\kappa$ stands for the screening parameter which serves here as a model parameter and is fixed at a value of 20 $\mathrm{nm}^{-1}$. 

Equation~\eqref{eq:EOM-e-pn} describes the coupling between electrons and phonons, which is the only process included that dissipates energy. We assume here a phonon bath with a cold Bose-Einstein distribution $N_{\bvec{q}} = (\exp(\hbar \omega_{\bvec{q}}/k_{\mathrm{B}} T_{\mathrm{eq}})-1)^{-1}$. The equilibrium temperature $T_{eq} = 70$ K is the bath temperature and $k_{\mathrm{B}}$ the Boltzmann constant. As in Ref.~\cite{Leckron2022}, we employ acoustic phonons in the long-wavelength limit, so that the phonon energy is linear, $\hbar \omega_{\bvec{q}} = c_{pn} q$, with a  sound velocity $c_{pn}$  of 10 nm ps$^{-1}$ and absolute value of the phonon momentum $q = |\bvec{q}|$. In this limit, the electron-phonon matrix element is given by $g_{\bvec{k} \nu , \bvec{k}_1 \nu_1} = D_{\mathrm{eff}}\sqrt{q}\langle\bvec{k}|\bvec{k}_1\rangle \delta_{\nu, \nu_1}$, as we assume no explicit spin-orbit coupling. The effective rescaled deformation potential $D_{\mathrm{eff}} = 10$
meV $\sqrt{\mathrm{nm}}$ is a material parameter and will be treated as a model parameter here.

Equations~\eqref{eq:EOM-ex-ss1} and~\eqref{eq:EOM-ex-ss2} contain the EOMs for exchange scattering (ex) between sublattices A and B with the exchange-interaction matrix element given by $ W_{\bvec{k} \nu, \bvec{k'} \nu'}^{\lambda \lambda'}=J\langle \bvec{k}, \nu|\hat{s}_z|\bvec{k'} ,\nu'\rangle \cdot \langle \lambda|\hat{S}_z|\lambda'\rangle$.  

The interactions so far describe exchange scattering between the sublattices and energy relaxation processes with phonons as the states in our model are taken to be pure spin states. In order to describe spin relaxation from the electronic system to the lattice we include an 
Elliot-Yafet-like process, i.e., an effective spin-flip scattering process that results from the combination of a spin independent scattering transition and the presence of spin orbit coupling. As we focus here on the exchange-scattering processes, we treat the spin relaxation processes at the level of a relaxation time approximation
\begin{equation}
   \frac{\partial}{\partial t} n^{\nu}_{\bvec{k}}\Bigg|_{\mathrm{sr}}  =  -\frac{n_{\bvec{k}}^{\nu}-f_{\bvec{k}}^{\nu}}{\tau_{\text{sr}}}
    \label{eq:relax-time}
\end{equation}
which describes the relaxation towards an equilibrium electron distribution $f_{\bvec{k}}^{\nu}$ with a relaxation time of $\tau_{\text{sr}}$. The distribution $f_{\bvec{k}}^{\nu}$ here is a Fermi-Dirac distribution with the same temperature as the non-equilibrium electron distribution $n_{\bvec{k}}^\nu$. On the time scale of the relaxation time $\tau=200$ fs angular momentum is thus transferred to a bath. This approach captures important aspects of spin-flip scattering due to spin-orbit coupling (SOC)~\cite{Leckron2019, Vollmar}.    

Equations~\eqref{eq:EOM-e-e}-\eqref{eq:relax-time} collectively constitute the fundamental system of EOMs governing the dynamics of exchange-coupled sublattices. We evaluate the EOMs in Eqs.~\eqref{eq:EOM-e-e}-\eqref{eq:EOM-ex-ss2} in the $\Gamma \to 0$ limit, in which case energy conserving delta-distributions arise in the scattering terms and we neglect the corresponding energy renormalizations beyond the mean-field contributions that are included in the instantaneous single-particle energies. Reliable numerical results during the entire dynamics depend on this energy conservation, as otherwise the system's energy density will be impacted by numerical errors, which can lead to errors in the  magnetization dynamics on short and long time scales.

\tikzset{
	electron1/.style={thick, draw=white!50!black , postaction={decorate},decoration={markings,mark=at position .7 with {\arrow[white!50!black]{triangle 45}}}},	
    electron2/.style={thick, draw=black , postaction={decorate},decoration={markings,mark=at position .7 with {\arrow[black]{triangle 45}}}},
	interaction/.style={thick,  draw=red, decoration={coil,aspect=0, segment length=16pt}, decorate}}
\begin{figure}[t]
	\centering
	\begin{tikzpicture}[scale=1]	
		\newcommand{\base}{440pt}
		\newcommand{\xstarta}{0.025*\base};
		\newcommand{\ystart}{-0.1*\base};
		\newcommand{\dx}{0.1*\base};
		\newcommand{\dy}{0.05*\base};	
		\newcommand{\xstartb}{\xstarta+2*\dx}
		\coordinate[] (ea1)	at (\xstarta-\dx,\ystart+\dy);
		\coordinate[] (xa)	at (\xstarta    ,\ystart);
		\coordinate[] (ea2)at (\xstarta-\dx,\ystart-\dy);
		\coordinate[] (eb1)	at (\xstartb+\dx, \ystart+\dy);
		\coordinate[] (xb)	at (\xstartb    , \ystart);
		\coordinate[] (eb2)	at (\xstartb+\dx, \ystart-\dy);
		\draw[electron1] (ea1) -- node[label=above:$\left|\mathbf{k}\mathrm{,}\nu\right\rangle$,yshift=5pt] {}(xa);
		\draw[electron2] (xa) -- node[label=above:$\left|\mathbf{k+q}\mathrm{,}\nu\right\rangle$, xshift=-15pt, yshift=-5pt] {}(ea2);
		\draw[interaction] (xa) -- node[label=below:$V_q$] {}(xb);
		\draw[electron1] (eb1) -- node[label=above:$\left|\mathbf{l}+\mathbf{q}\mathrm{,}\nu'\right\rangle$, yshift=5pt] {}(xb);
		\draw[electron2] (xb) -- node[label=above:$\left|\mathbf{l}\mathrm{,}\nu'\right\rangle$, xshift=15pt, yshift = -5pt] {}(eb2);
	\end{tikzpicture}	
	\caption{Illustration of the electron-electron Coulomb scattering process between electronic states in sublattice A: $(\mathbf{k},\nu), (\mathbf{l}+\mathbf{q},\nu') \to (\mathbf{k+q},\nu), (\mathbf{l},\nu')$. The band indices label the states in sublattice A. }
	\label{fig:e-e-sketch}
\end{figure}
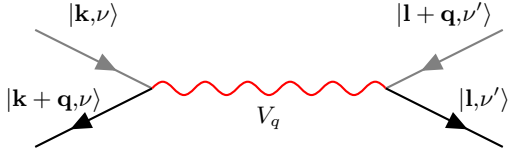

\subsection{Equilibrium band structure and model excitation\label{sec:model}}

We obtain the equilibrium band configuration and corresponding distributions for the states self consistently. Throughout this paper we assume an equilibrium temperature of $T_{eq}=70$\, K. This choice lies significantly below the Curie temperature of $T_{C}=$ 274 K for the parameters of the model, see the caption of Fig.~\ref{fig:bands}.

We will consider two distinct instantaneous excitation conditions. One is intended to mimic the direct effect of an optical field that creates strongly $k$-dependent excited electron distributions. The other one is an indirect mechanism that creates hot electrons.  The indirect excitation is intended to capture aspects of the hot-carrier injection realized in Ref.~\cite{Schmising}. Regarding the influence on the magnetization dynamics the two different excitation conditions may be called primary and secondary effects, respectively, with respect to the influence of external fields: primary, because the $k$ dependent nature of the excitation imprinted by the optical field is important, secondary, because the deposited energy plays a role which has no direct relation to any optical transitions in the system. We assume that both excitation mechanisms affect only the itinerant electrons in sublattice A, which have a considerably larger $k$ dispersion, i.e., band width, than the flat bands in sublattice B.

\begin{figure}[t]
   \centering
    \includegraphics[width=\columnwidth]{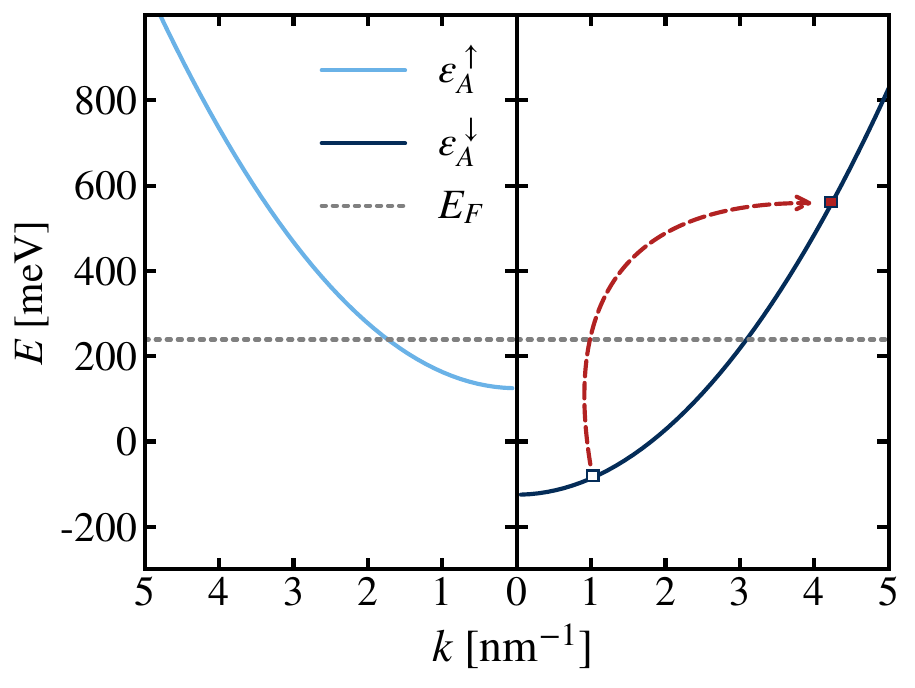}
	\caption{Schematic illustration of the instantaneous k-dependent excitation process that approximates the effect of an ultrashort pulse creating an excitation in the majority spin channel of sublattice A. The gray dashed line refers to the Fermi energy level. The parameters $U_{\mathrm{eff}}$ and $J$ are the same as in Fig.~\protect\ref{fig:bands}. The zero of the energy axis is the band bottom without spin splitting.
	\label{fig:peaked_excitation}}
\end{figure}

We first describe a direct $k$ dependent excitation of carriers in sublattice A, as shown in Fig.~\ref{fig:peaked_excitation}. We model this excitation process by a Gaussian-shaped electron distribution above the Fermi edge in the majority band of sublattice A. This is the idealized case of a majority-electron excitation  so that the influence of scattering processes can be seen most directly. Note that this process conserves the density and does not alter the magnetization. In a more realistic multi-band model, the character of the excitation can be majority or minority-electron dominated and depends on the photon energy Ref.~\cite{Stiehl}.  

The second excitation process considered in the following is modeled by the instantaneous creation of a hot electron distribution~\cite{Qaiumzadeh2013,Leckron2022}. It involves the instantaneous replacement of the thermal the Fermi-Dirac (FD) distribution in sublattice A by a hot electron distribution in the form of a FD distribution with an elevated temperature $T_{\mathrm{H}} \gg T_{\mathrm{eq}}$.
\begin{equation}
	n^{\nu}_{\bvec{k}}= \frac{1}{e^{(
\epsilon^{\nu}_{\bvec{k}}-\mu)/(k_{B}T_{\mathrm{H}})}+1  }   ,
	\label{eq:Fermi-Dirac}
\end{equation}
Here, $\mu$ represents a self-consistently calculated chemical potential, $k_{B}$ is Boltzmann's constant, $T_{\mathrm{H}}$ is the excitation temperature. This excitation mechanism itself also preserves the magnetization, so that the subsequent magnetization dynamics are exclusively due to scattering processes.

\section{Results\label{sec:result}}

In this section, we present the electronic dynamics computed with Eqs.~\eqref{eq:EOM-e-e}--\eqref{eq:relax-time} for the two excitation scenarios described above.

\subsection{Electron dynamics for $\mathbf{k}$-dependent excitation}

We first consider a $k$-dependent, i.e., energy selective excitation, as it is typically created by optical transitions. We model this by an instantaneous change in distribution functions $\delta n_\bvec{k}^{\nu}$ and plot in the following, the occupation difference 
\begin{equation}
    \delta n_\bvec{k}^{\nu}(t) = n_\bvec{k}^{\nu}(t) - f_\bvec{k}^{{\nu}(eq)},
\end{equation}
between the time-dependent carrier distribution $n_\bvec{k}^{\nu}(t)$ and the stationary equilibrium distribution function $f_\bvec{k}^{{\nu}(eq)}$, see Eq.~\eqref{eq:Fermi-Dirac}, at the temperature $T_{\mathrm{eq}} = 70\;\mathrm{K}$.
\begin{figure}[t!]
    \centering
  \includegraphics[width=\columnwidth]{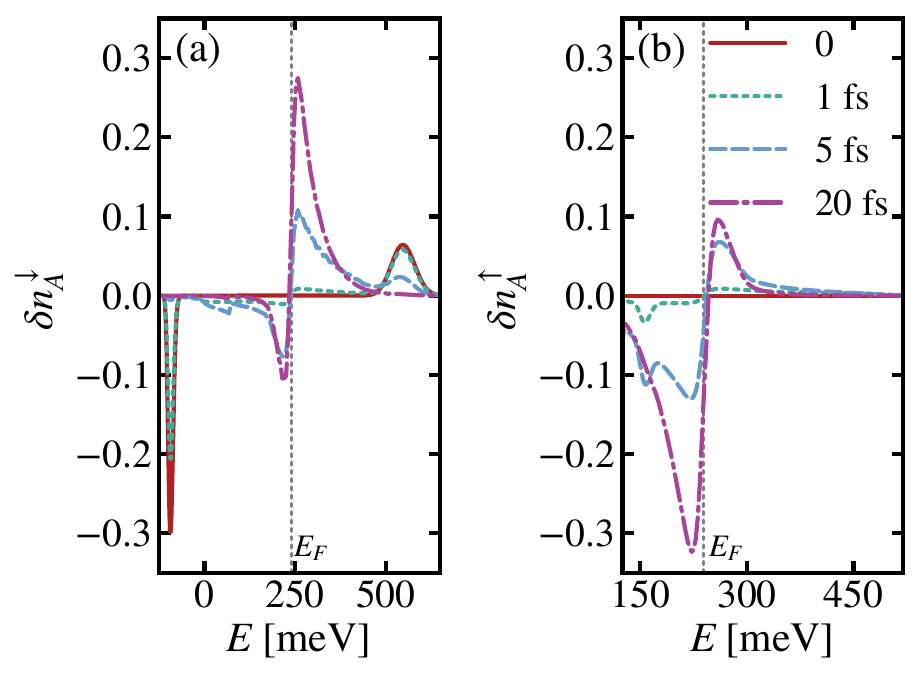}
	\caption{Snapshots of electronic occupation change with respect to a cold Fermi-Dirac distribution at $T_{\mathrm{eq}} = 70 K$ in the majority $\delta n_{A}^{\downarrow}$ (a) and minority band $\delta n_{A}^{\uparrow}$ (b) after optical excitation including e-e, e-pn, and ex scattering contributions. The parameters $U_{\mathrm{eff}}$ and $J$ are the same as in Fig.~\ref{fig:bands} and $E_F$ denotes the Fermi energy level.}
\label{fig:peaked_excitation_ee_pn_ex}	
\end{figure}
  
\begin{figure}[t!]
    \centering   \includegraphics[width=\columnwidth]{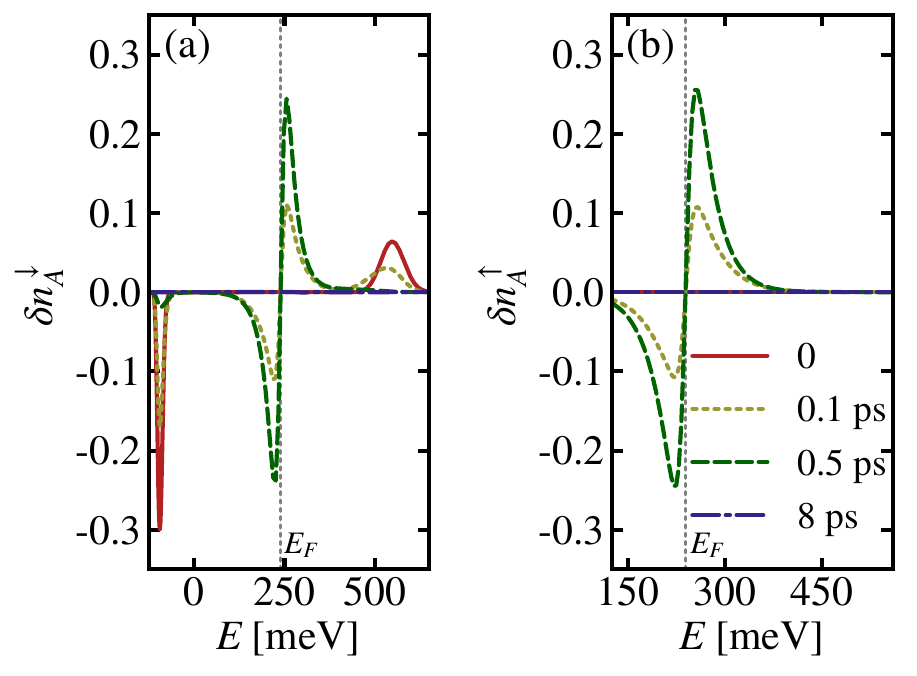}
    \caption{Same as Fig.~\ref{fig:peaked_excitation_ee_pn_ex}, except that e-e scattering is excluded.}
\label{fig:peaked_excitation_pn_ex}
\end{figure}

Figure~\ref{fig:peaked_excitation_ee_pn_ex} shows the occupation change in the minority (a) and majority band (b) for different times including e-e, e-pn, and exchange scattering contributions. Figures~\ref{fig:peaked_excitation_pn_ex}(a) and~(b) display the same quantity computed including only e-pn, effective EY-spin-flips and ex scattering, respectively. 

The excitation process at $t=0$ leads to a structure with a decrease in electron occupation around $E=-93\;\mathrm{meV}$ and an increase around $E=548\;\mathrm{meV}$. This idealized excitation is chosen such that the electron density before and after excitation is conserved. Since we use an effective 2D model, the peaks have a different height as well as width in energy because of the volume element scaling with $k$ and the quadratic dispersion of the electrons in sublattice A. Initially, the electrons in minority  band are still in equilibrium. 

An obvious difference between Figs.~\ref{fig:peaked_excitation_ee_pn_ex} and~\ref{fig:density_pn_ex} is the time scale of the dynamics, which is much faster with e-e scattering included and results in a quick washing out of the initial peaks in Fig.~\ref{fig:peaked_excitation_ee_pn_ex}~(a). 

We focus first on the timescale in Fig.~\ref{fig:peaked_excitation_ee_pn_ex} after 1 fs when e-pn scattering does not yet play a role. At these early times, e-e scattering processes redistribute electrons in the majority band over the whole $k$-space/energy range towards a quasi-equilibrium distribution, i.e., a ``hot'' FD distribution. It is this spin-conserving redistribution in $k$-space that opens up scattering phase space for exchange scattering processes such as $|\uparrow\rangle \rightarrow |\downarrow\rangle$ in sublattice A accompanied by $|-\rangle \rightarrow |0\rangle$ in sublattice B. 

\begin{figure}[t!]
    \centering
    \includegraphics[width=\columnwidth]{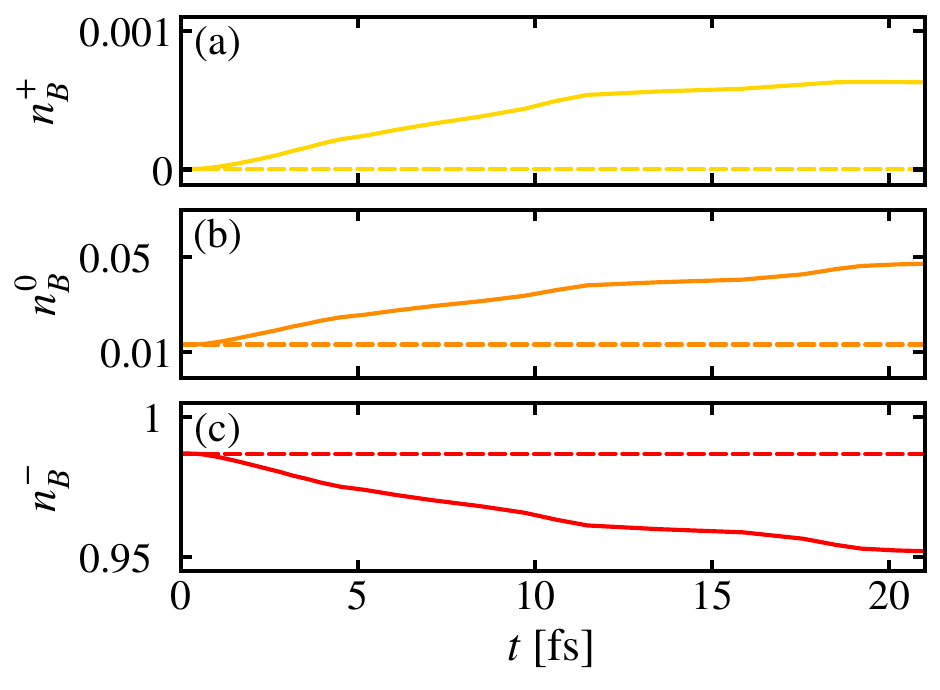}
    \caption{Time-dependent occupation of bands $n_{B}^{+}$, $n_{B}^{0}$ and $n_{B}^{-}$ in sublattice B after peaked excitation, where solid lines refer to the case in Fig.~\ref{fig:peaked_excitation_ee_pn_ex} (e-e, e-pn, EY spin-flip and ex) and the dotted lines to the case in Fig.~\ref{fig:peaked_excitation_pn_ex} (e-pn, EY spin-flip and ex). The parameters $U_{\mathrm{eff}}$ and $J$ are the same as in Fig.~\ref{fig:bands}.}
\label{fig:peaked_excitation_lokal}
\end{figure}

The corresponding dynamics of sublattice B for the different combination of scattering processes, i.e., (e-e, e-pn and ex) vs. (e-pn and ex) are illustrated in  Fig.~\ref{fig:peaked_excitation_lokal}. The dynamical distributions $n_{B}^{+}$, $n_{B}^{0}$ and $n_{B}^{-}$ are shown as solid and dotted lines for the respective combinations of scattering mechanisms.  
Since the bands in sublattice B are flat, there is only a single fixed energy, i.e., the band gap size in sublattice B, $\Delta E_{B}^{\lambda}$, for which these kinds of transitions (e.g., $|-\rangle \rightarrow |0\rangle$ and $|0\rangle \rightarrow |+\rangle$) are possible. At very early times in Fig.~\ref{fig:peaked_excitation_lokal}~(a) there is a decrease in the occupations $n_{B}^{-}$ of the lowest band, which is accompanied, cf.~Figs.~\ref{fig:peaked_excitation_lokal}~(b) and~(c), by an increase in the occupations  $n_{B}^{0}$ and $n_{B}^{+}$ of bands at higher energies.      
Referring to Fig.~\ref{fig:peaked_excitation_ee_pn_ex}~(b), it is these transitions that cause a characteristic ``dip'' in the minority bands below 200 meV. The electrons at this energy in the minority band of sublattice A are filling the states from which electrons were redistributed to higher energies 
in the exciation process. due to their energy difference which fits $\Delta E_{B}^{\lambda}$, thereby transferring energy from sublattice A to sublattice B. The dip in the minority band gets constantly redistributed to the rest of the $k$-space in the minority band and therefore washed out. 

\begin{figure}[t!]
    \centering
    \includegraphics[width=\columnwidth]{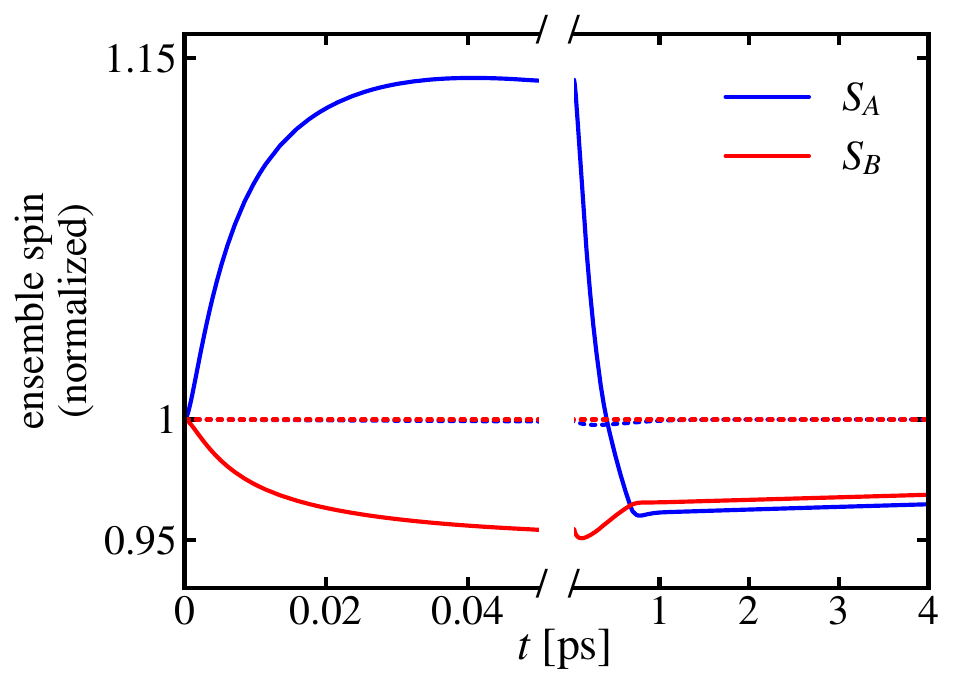} 
    \caption{Comparison of the normalized ensemble spin dynamics for both sublattices A ($S_A$) and B ($S_B$) under two distinct scattering scenarios: solid lines for e-e, e-pn, EY spin-flip and exchange scattering and dotted lines only e-pn, EY spin-flip and exchange scattering for $k$-dependent excitation.
    The parameters $U_{\mathrm{eff}}$ and $J$ are the same as in Figs.~\ref{fig:bands}. }   
    \label{fig:peaked_excitation_spin}
\end{figure}

The source of the dip in the minority band becomes clearer for the 5 fs line in the majority band, visible as a ``kink'' for energies close to 100 meV in Fig.~\ref{fig:peaked_excitation_ee_pn_ex}~(a). The holes are transferred to the minority band starting at the high-energy side of the kink. At 5 fs, the initial peak in the majority band below the Fermi edge, which was created by the excitation, has almost completely disappeared. In contrast, the excitation peak above the Fermi edge is still noticeable, as there are only few energetically allowed exchange scattering process in which these electrons at higher energies could take part. 
At 20 fs the electron system has reached a quasi-equilibrium state characterized by an increase in electron density in the majority band and a decrease in the minority band, as can be seen by the form of the bipolar shape of the distributions at 20 fs in Fig.~\ref{fig:peaked_excitation_ee_pn_ex}. In a hot FD distribution, the lobes to the left and right of the Fermi edge should be equal in size but different in sign, as shown for the dynamics without e-e scattering below, see the 0.5 ps lines in Fig.~\ref{fig:peaked_excitation_pn_ex}. However, the 20 fs line in Fig.~\ref{fig:peaked_excitation_ee_pn_ex}~(a) is asymmetric with a larger feature above $E_F$. Stated differently, the majority band has more electrons than in an actual FD distributed equilibrium distribution, and a larger portion below $E_F$ for the minority band, indicating fewer electrons than in an equilibrium situation. This non-equilibrium state also involves a higher occupation of the $n_{B}^{0}$ band for sublattice B (see Fig.~\ref{fig:peaked_excitation_lokal}~(a) at t=20 fs). This has consequences for the ensemble spin dynamics as discussed later on in Fig.~\ref{fig:peaked_excitation_spin}.

When e-e scattering is neglected, as shown in Fig.~\ref{fig:peaked_excitation_pn_ex}, no such imbalance between the minority and majority bands of sublattice A results. The time scale of the dynamics is much longer due to the weaker e-pn scattering, which now governs both the thermal distributions, i.e., the formation of a (hot) FD distribution, and also cooling. The excitation peaks now disappear on a picosecond timescale and no exchange-induced `dips' are visible in the minority band, see Fig.~\ref{fig:peaked_excitation_pn_ex}~(b). The redistribution of the initial excitation happens slowly so that a sizeable scattering phase space for a pronounced energy dependent signature of the exchange scattering as in Fig.~\ref{fig:peaked_excitation_ee_pn_ex}~(a) is never opened up. This leads to only a very subtle change in occupation of the $n_{B}^{0}$ band in sublattice B.  Moreover, as it is shown in Fig.~\ref{fig:peaked_excitation_lokal}~(dotted lines), there is no considerable change in the occupation of the bands of sublattice B in the short timescales nor in the longer ones up to 8 ps. 

We next study the consequences of the microscopic dynamics for the change of the ensemble spin in sublattices A and B shown in Fig.~\ref{fig:peaked_excitation_spin} normalized to its initial equilibrium value.
There, the spin dynamics with and without e-e scattering can be directly compared. For the case with e-e scattering, we see an ultrafast \emph{increase} in the ensemble spin of sublattice A accompanied by an ultrafast decrease in the ensemble spin of sublattice B in roughly 40 fs. This is the result of the interplay of the ultrafast electron redistribution of the excitation due to e-e scattering and the corresponding opening of phase space for the exchange scattering, which transfers electrons from the $|\uparrow\rangle$ to the $|\downarrow\rangle$ band in sublattice A accompanied by a promotion of electrons $|-\rangle \rightarrow |0\rangle$ in sublattice B (see Fig.~\ref{fig:peaked_excitation_lokal}~(a-b)). For a longer timescale, a typical demagnetization on a 100 fs scale and a remagnetization on a much longer ps scale are also shown. 

Without e-e scattering, by contrast, the ensemble spin dynamics in sublattice A are not only much weaker, but we also do not see an increase in ensemble spin at all. Due to the extremely limited scattering phase space discussed above the dynamics of the ensemble spin of sublattice B are not yet visible on the timescale used in Fig.~\ref{fig:peaked_excitation_spin}.   

The discussion so far has shown electron-electron scattering plays an important role in the magnetization dynamics when the exchange coupled two-sublattice system experiences a $k$-dependent electronic excitation far from the Fermi edge, as is the case for ultrashort optical pulses. This shows that even when the spin-dependent effect is due solely to the exchange interaction that flips the spin between the subsystems, spin conserving e-e scattering yields an important contribution to the ultrafast dynamics if its scattering phase space is activated by the excitation. The combination of the e-e scattering with the exchange scattering also leads to specific microscopic electronic dynamics, i.e., occupation changes in energy regions that would not occur for either scattering mechanism alone. These fingerprints of the interplay of the scattering mechanisms may be accessible using present-day experimental techniques. 

\subsection{Electron dynamics for hot electron excitation}

In this subsection we consider an excitation with hot electrons. This is useful for two scenarios: It can correspond to an injection of hot electrons or one can use this excitation condition to study the effect of the deposited energy alone. In the latter case one takes the results of the previous subsection into account and argues that the hot-electron quasi-equilibrium will be established by the e-e scattering processes on ultrafast timescales. The computed dynamics would then be only a good approximation to the full calculation for timescales longer than those needed by the e-e scattering to establish a quasi-equilibrium. 

In the following, we choose for initial condition an excitation with hot electrons at an effective temperature of $T_{H} = 1500$ K ($T_{H}\gg T_{eq}$), but with the same magnetization as in total equilibrium. 

\begin{figure}[t!]
   \centering
	\includegraphics[width=\columnwidth]{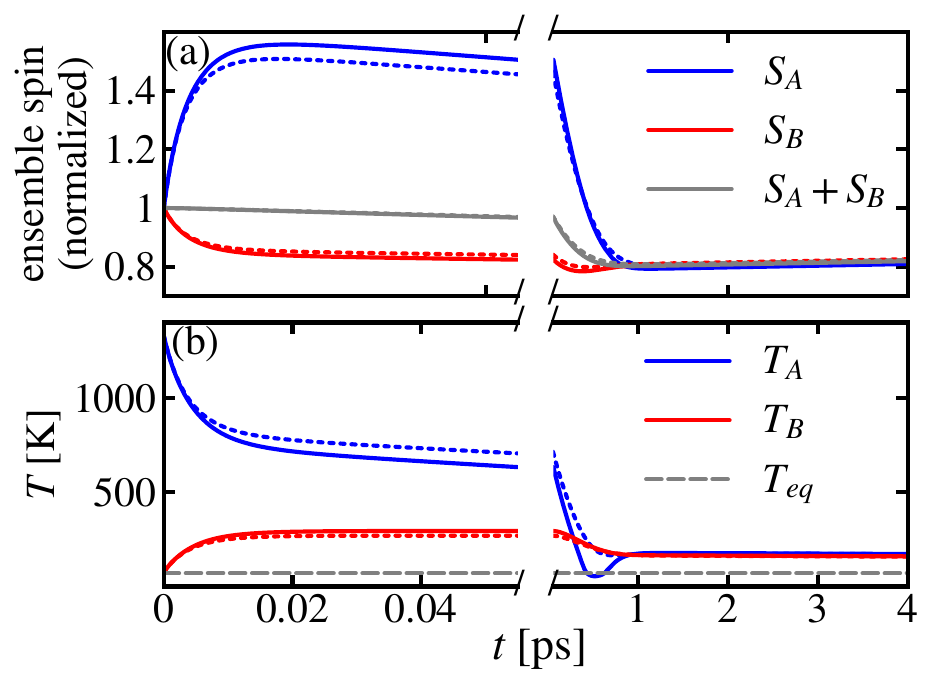}
	\caption{Ultrafast dynamics of the normalized ensemble spin expectation value (a), temperatures (b) in both sublattice A and B for the two different scattering mechanism scenarios  (i.e., solid lines (e-e,e-pn and ex) and dotted lines (e-pn and ex)). Here, sublattice A is excited at $T_{H}$. The gray dashed line in (b) refers to phonon bath temperature at $T_{eq}$. The parameters $U_{\mathrm{eff}}$ and J here are as same as Fig.~\ref{fig:bands}}
	\label{fig:spin_temp_un100_tx1500}	
\end{figure}

Figure~\ref{fig:spin_temp_un100_tx1500}(a) shows the normalized ensemble spin and Fig.~\ref{fig:spin_temp_un100_tx1500}(b) the effective temperature in both sublattices. These curves show ensemble averaged quantities, which are obtained from the dynamical solution of the microscopic equations of motion as in the previous subsection, but for the hot-electron excitation. The effective temperature shown in Fig.~\ref{fig:spin_temp_un100_tx1500}(b) is not a thermodynamical quantity but it is determined by a fitting procedure as the temperature of an equilibrium distribution with the same energy density as the non-equilibrium distribution evolving according to Eqs.~\eqref{eq:EOM-e-e}--\eqref{eq:relax-time}. In Figs.~\ref{fig:spin_temp_un100_tx1500}(a) and (b) solid lines correspond to calculations including e-e, e-pn, EY spin-flip and exchange scattering mechanisms and the dashed lines to calculations in which the e-e scattering mechanism is switched off. These results indicate that during the initial femtoseconds of the dynamics, the total spin of sublattice A increases more rapidly at around $t=15$ fs accompanied by a decrease in temperature when considering spin-dependent electron-electron scattering. In contrast, the magnetization of the sublattice B decreases on ultrashort timescales while it heats up more rapidly in the presence of electron-electron scattering. 

\begin{figure}[t!]
   \centering
\includegraphics[width=\columnwidth]{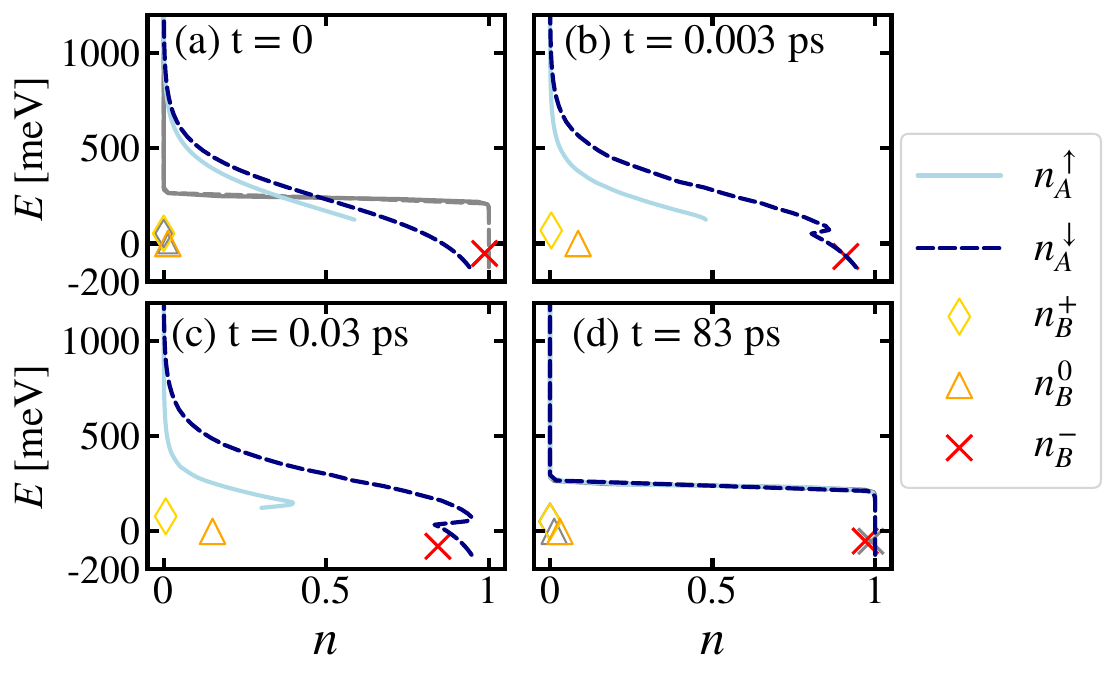}
	\caption{Energy band resolved distribution dynamics of bands $n_{A}^{\uparrow}$ and $n_{A}^{\downarrow}$ in sublattice A and bands $n_{B}^{+}$, $n_{B}^{0}$ and $n_{B}^{-}$ in sublattice B. The color/symbol code in (a) and (d) denotes the equilibrium distribution. The scattering mechanisms involved in the dynamics are e-pn, EY spin-flip, and ex scattering.
 The parameters $U_{\mathrm{eff}}$ and $J$ are as same as Fig.~\protect\ref{fig:spin_temp_un100_tx1500}.}
	\label{fig:density_pn_ex}	
\end{figure}

\begin{figure}[t!]
   \centering
	\includegraphics[width=\columnwidth]{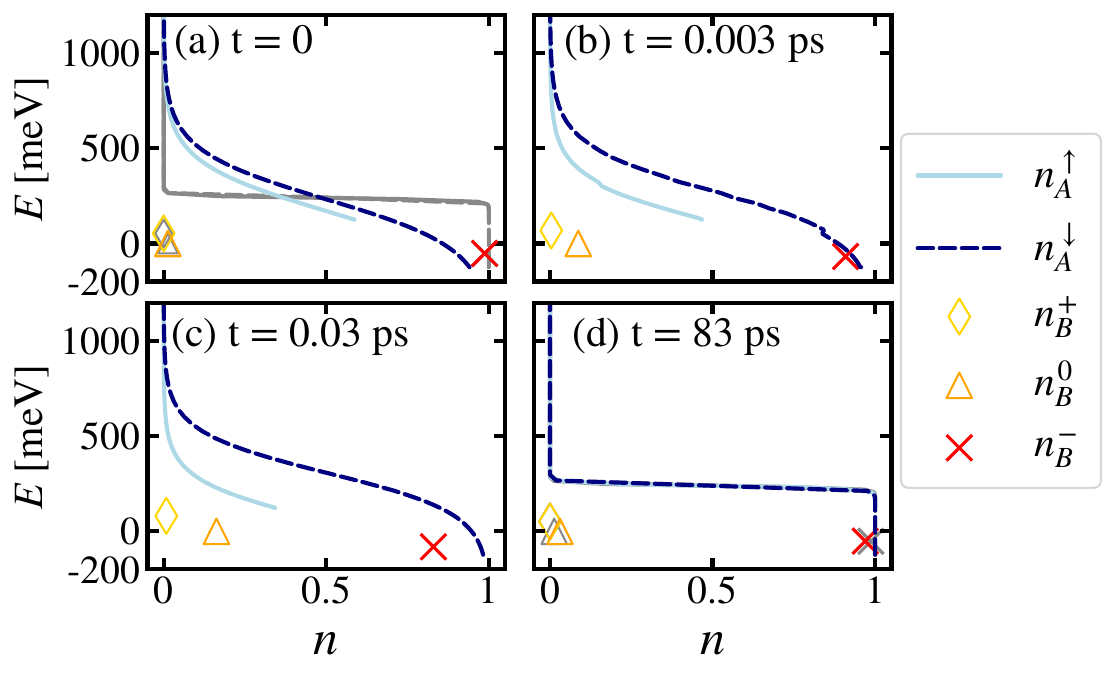}
	\caption{Same as Fig.~\ref{fig:density_pn_ex}, but e-e scattering is included.}
\label{fig:density_ee_epn_ex}	
\end{figure}

The microscopic picture of the dynamics as snapshots of electronic distributions for both scattering scenarios is illustrated in Fig.~\ref{fig:density_pn_ex} without electron-electron scattering  and Fig.~\ref{fig:density_ee_epn_ex} including electron-electron scattering. Figs.~\ref{fig:density_pn_ex}--\ref{fig:density_ee_epn_ex} (a) show electronic distributions of both sublattices after the instantaneous hot electron injection process at $t=0$ in sublattice A with their equilibrium distribution in gray color/symbol code. The instantaneous excitation processes is such that the distribution of sublattice A in both Fig.~\ref{fig:density_pn_ex} and Fig.~\ref{fig:density_ee_epn_ex} at $t = 0$ is a ``hotter'' FD distribution while the distribution of the bands in sublattice~B is unaffected. After the next few femtoseconds, which is the intrinsic timescale of the exchange interaction, Figs.~\ref{fig:density_pn_ex}-\ref{fig:density_ee_epn_ex} (b) shows that the electronic distribution in both sublattices have changed due to exchange scattering in which all electrons possessing energy in excess of $\epsilon_{\bvec{k}}^{\uparrow}-\epsilon_{\bvec{k}'}^{\downarrow} = E_0 - E_-$ scatter within sublattice A $\uparrow\rightarrow \downarrow$ and sublattice B $|-\rangle \rightarrow |0\rangle$.

A consequence of the exchange scattering mechanism is that some electrons in sublattice A scatter from the minority band to the majority band. These electrons are unable to reach the smaller energy range located at the bottom of the majority band and therefore in Figs.~\ref{fig:density_pn_ex}(b-c) after around 3 fs a kink develops within the electronic distribution of the majority band $|\downarrow\rangle$ which points out the absence of the electrons in the lower energy region of the majority band. 
Within the minority ($\uparrow$) band of sublattice A in Fig.~\ref{fig:density_pn_ex}(b-c) a kink develops, too, but it results from the modified instantaneous quasiparticle band structure where the band splitting $\Delta$ of sublattice A depends on the ensemble spin expectation value of both sublattices, cf.~Eqs.~\eqref{eq:itinen}-\eqref{eq:itindelta}. On this timescale, there is an increase in the ensemble spin expectation value of sublattice A and a corresponding decrease in the ensemble spin expectation value of sublattice B, see Fig.~\ref{fig:spin_temp_un100_tx1500}(a), which increases the band splitting in sublattice A and thus shifts the minority band to higher energies. 


Figure~\ref{fig:density_ee_epn_ex} presents the corresponding results obtained including electron-electron scattering contributions are shown on the same timescales. In this case, the intrinsic timescales of  spin-conserving electron-electron scattering in sublattice A are similar to those of the exchange scattering processes between sublattices A and B. Consequently, electrons within both the majority $|\downarrow\rangle$ and minority  $|\uparrow\rangle$ bands of sublattice A are driven towards a quasi-equilibrium by a scattering process that redistributes electrons in $k$-space with the constraint that the kinetic energy in sublattice A and the spin expectation value in the individual A bands be constant. Due to the phase space opened up by the excitation, the small kinks observed at $t=3$ fs in Fig.~\ref{fig:density_ee_epn_ex}(b) are completely washed out within the initial 30 fs of the electron dynamics depicted in Fig.~\ref{fig:density_ee_epn_ex}(c).  
 

We next investigate the dynamics on longer timescales for both scattering scenarios. We start with the right side of Fig.~\ref{fig:spin_temp_un100_tx1500} which covers the dynamics up to 4 ps. All the scattering mechanisms included in each scattering scenario work together on this timescale. Fig.~\ref{fig:spin_temp_un100_tx1500}(a) shows that, in both scattering scenarios, the normalized magnetization of sublattices A and B and their normalized total ensemble spin $S_A + S_B$ reach $80\%$ of their equilibrium values at around $t=1$ ps. The effective electronic temperatures of sublattice A and B in Fig.~\ref{fig:spin_temp_un100_tx1500}(b) almost converge to a quasi-equilibrium value at the same time. Only the carrier dynamics including electron-electron scattering exhibits a pronounced dip in the effective temperature of the itinerant electrons in sublattice A before 1\,ps.

Once the magnetization and effective temperatures of the sublattices have reached a quasi-equilibrium they converge to their equlibrium values characterized by a common temperature  $T_{eq}$. This slow relaxation process relies on the transfer of energy from sublattice A into the phonon bath characterized by the time scale of Eq.~\eqref{eq:relax-time}. Sublattice B is only indirectly cooled via the exchange interaction with the sublattice A. On this timescale there is no difference between the dynamics with and without the spin-conserving electron-electron scattering.   

For completeness, the return to equilibrium on the timescale of several 10 ps is also shown in the snapshot of the distribution functions at 83 ps in Figs.~\ref{fig:density_pn_ex}(d) and \ref{fig:density_ee_epn_ex}(d). On this long time scale the electron-electron scattering has no visible influence and an effective description in terms of local temperatures and chemical potentials is applicable.

We now investigate the influence of the spin-conserving electron-electron scattering mechanism on the spin dynamics in dependence of some of the parameters that define our model. An increase of the Stoner parameter leads to a more robust ferromagnetic behavior of subsystem A. In a localized spin model this would correspond to increasing the intra-sublattice~A exchange constant. In the itinerant electron picture this leads to a band structure where electrons predominantly occupation of majority spin states due to the larger spin splitting for the same carrier density, according to Eqs.~\eqref{eq:itinen}--\eqref{eq:itindelta} this results also in larger gaps in sublattice B. We choose here a Stoner parameter to $U_{\mathrm{eff}} = -330$ meV. 

\begin{figure}[t]
\centering
\includegraphics[width=\columnwidth]{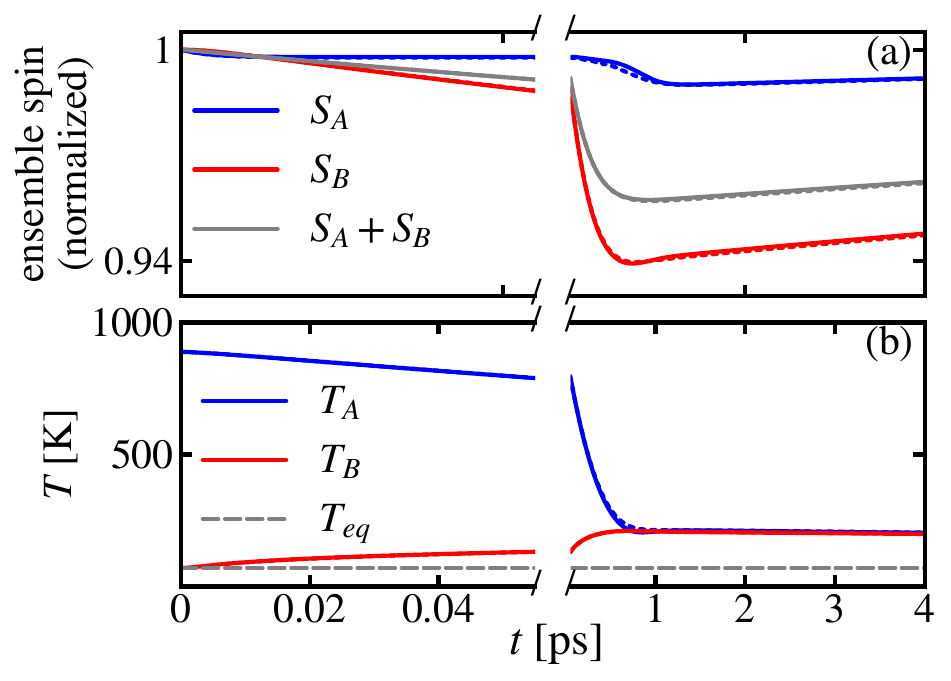}
	\caption{
    Same as Fig.~\ref{fig:spin_temp_un100_tx1500} for $U=-330$ meV.}	\label{fig:spin_temp_un330_tx1500}
\end{figure}

Figure~\ref{fig:spin_temp_un330_tx1500}(a) depicts the dynamics of the normalized ensemble spin in both sublattices as well as their total ensemble spin. The corresponding effective temperature of the sublattices is shown in Fig.~\ref{fig:spin_temp_un330_tx1500}(b). The solid lines show the dynamics on ultrashort and short timescales including all interaction mechanisms while the dotted lines are the results without spin-conserving electron-electron scattering.
Compared to Fig.~\ref{fig:spin_temp_un100_tx1500}, the change of the ensemble spin is much weaker and does not show the redistribution of electrons due to exchange scattering on ultrafast timescales which leads to the initial increase of the ensemble spin in sublattice A with a corresponding decrease of the ensemble spin in sublattice B. This is a result of the larger band gap, i.e., there are not enough electrons available for exchange scattering transitions in the minority band in equilibrium so the excitation via a spin-conserving hot FD distribution cannot lead to electrons from the minority band scattering to the majority band. 

In contrast, the demagnetization observed here is due to spin-flip processes as one would also observe in an elemental metallic ferromagnet, as discussed for the case without spin-conserving e-e scattering in Ref.~\cite{Leckron2022}. We therefore foucs on the influence of the additional e-e scattering. It turns out, however, that this contribution has only a very small effect on the overall dynamics. If the system is excited in a quasi-equilibrium, either because of the injection mechanism, or because a quasi-equilibrium at an an elevated temperature has been established by scattering processes, the momentum resolved spin-dependent intraband electron-electron scattering is mostly negligible as the system never evolves far away from a quasi-equilibrium. 

\section{Conclusion\label{sec:conclusion}}

We investigated the electron dynamics in a two-sublattice model following an idealized instantaneous excitation processes with particular focus on the influence of momentum-resolved electron-electron scattering treated at the level of Boltzmann scattering integrals. In the framework of a model bandstructure with time-dependent quasiparticle states we included exchange scattering, spin conserving intraband electron-electron scattering, energy dissipating electron-phonon scattering and an effective spin-flip relaxation processes. We computed the carrier and ensemble spin dynamics for a direct excitation scenario which creates initial distributions far from equilibrium and a hot-carrier injection scenario, which can be characterized by a deposited energy. 
For the direct excitation scenario, spin conserving electron-electron scattering have a noticable impact on the microscopic distribution function dynamics as well as the ensemble spin dynamics. In the case of the hot-electron excitation with a small ferromagnetic splitting in sublattice A, the influence is less pronounced, albeit still visible, especially the ultrafast disappearance of the ``kink'' in the electron distribution when including electron-electron scattering. For a pronounced ferromagnetic splitting in one of the sublattices,  the influence of exchange scattering processes and the momentum-resolved features driven by this process become weaker. Consequently, the influence of electron-electron scattering which evens out "kinks" in the electronic distributions becomes hardly noticable.

We conclude that electron-electron scattering processes in two-sublattice systems need to be taken into account if far-from-equilibrium electronic distributions are excited, but even for hot-electron excitation scenarios where a quasi-equilibrium picture should be applicable, exchange scattering processes may drive the system away from quasi-equilibrium which is then restored by electron-electron scattering processes. Electron-electron scattering processes lead to fingerprints in the energy-resolved electron distributions on ultrafast timescales in two-sublattice magnets that likely cannot be described by localized-spin models and may be accessible by present-day experimental techniques.

\begin{acknowledgments}
Funded by the Deutsche Forschungsgemeinschaft
(DFG, German Research Foundation) – TRR 173 – 268565370 Spin+X (Project No. A08). We acknowledge a CPU time grant at the high-performance cluster Elwetritsch, which is part of the Alliance of High Performance Computing Rheinland-Pfalz (AHRP).
\end{acknowledgments}

\section*{Data Availability Statement}
The data supporting this study's findings are available from the corresponding author upon reasonable request.


%

\end{document}